\documentclass[aps,twocolumn,secnumarabic,nobalancelastpage,superscriptaddress,amsmath,amssymb,nofootinbib,reprint,prb]{revtex4-2}
\usepackage{graphicx}
\usepackage[version=3]{mhchem}
\usepackage{graphicx}
\usepackage{dcolumn}
\usepackage{bm}
\usepackage{hyperref}
\hypersetup{
    colorlinks=true,
    linkcolor=blue,
    filecolor=blue,
    urlcolor=blue,
   citecolor=blue,
}
\usepackage{enumitem}
\usepackage{siunitx}
\usepackage{inputenc}
\usepackage{braket}

\begin{document}

\title{Spin-Polarized Majorana Zero Modes in Proximitized Superconducting Penta-Silicene Nanoribbons}

\author{R. C. Bento Ribeiro}
\affiliation{Centro Brasileiro de Pesquisas F\'{i}sicas, Rua Dr. Xavier Sigaud, 150, Urca 22290-180, Rio de Janeiro, RJ, Brazil}

\author{J. H. Correa}
\affiliation{Universidad Tecnológica del Perú, Nathalio Sánchez, 125, 15046, Lima, Perú}
\affiliation{AGH University of Krakow, Academic Centre for Materials and Nanotechnology, al. A. Mickiewicza 30, 30-059 Krakow, Poland}

\author{L. S. Ricco}
\affiliation{Science Institute, University of Iceland, Dunhagi-3, IS-107,Reykjavik, Iceland}

\author{I. A. Shelykh}
\affiliation{Science Institute, University of Iceland, Dunhagi-3, IS-107,Reykjavik, Iceland}
\affiliation{Russian Quantum Center, Skolkovo IC, Bolshoy Bulvar 30 bld. 1, Moscow 121205, Russia}

\author{Mucio A. Continentino}
\affiliation{Centro Brasileiro de Pesquisas F\'{i}sicas, Rua Dr. Xavier Sigaud, 150, Urca 22290-180, Rio de Janeiro, RJ, Brazil}

\author{A. C. Seridonio}
\affiliation{School of Engineering, Department of Physics and Chemistry, S\~ao Paulo State University (UNESP), 15385-000 Ilha Solteira-SP, Brazil}

\author{M. Minissale}
\affiliation{Aix-Marseille Universit\'{e}, CNRS, PIIM UMR 7345, 13397 Marseille Cedex, France}

\author{G. Le Lay}
\affiliation{Aix-Marseille Universit\'{e}, CNRS, PIIM UMR 7345, 13397 Marseille Cedex, France}

\author{M. S. Figueira}
\affiliation{Instituto de F\'{i}sica, Universidade Federal Fluminense, Av. Litor\^anea s/N, CEP: 24210-340, Niter\'oi, RJ, Brasil}

\date{\today}
\begin{abstract}
{We theoretically investigate the possibility of obtaining Majorana zero modes (MZMs) in penta-silicene nanoribbons (p-SiNRs) with induced \textit{p}-wave superconductivity. The model explicitly considers an external magnetic field perpendicularly applied to the nanoribbon plane, as well as an extrinsic Rashba spin-orbit coupling (RSOC), in addition to the first nearest neighbor hopping term and \textit{p}-wave superconducting pairing. By analyzing the dispersion relation profiles, we observe the successive closing and reopening of the induced superconducting gap with a single spin component, indicating a spin-polarized topological phase transition (TPT). Correspondingly, the plots of the energy spectrum versus the chemical potential reveal the existence of zero-energy states with a preferential spin orientation characterized by nonoverlapping wave functions localized at opposite ends of the superconducting p-SiNRs. These findings strongly suggest the emergence of topologically protected, spin-polarized MZMs at the ends of the p-SiNRs with induced \textit{p}-wave superconducting pairing, which can be realized by proximitizing the nanoribbon with an \textit{s}-wave superconductor, such as lead. The proposal paves the way for silicene-based Majorana devices hosting multiple MZMs with a well-defined spin orientation, with possible applications in fault-tolerant quantum computing platforms and Majorana spintronics.}
\end{abstract}
\maketitle

\section{Introduction}

Ultra-scaling of nanoelectronic devices, beyond Moore’s law, still using the ubiquitous silicon technology, could come from silicene \cite{vogt2012silicene,Fleurence2012,Feng2012}, the first silicon-based graphene-like artificial two-dimensional (2D) quantum material, which further engendered the Xenes family \cite{davila2022silicene}, and which was used to fabricate an atom-thin channel in a field effect transistor \cite{tao2015silicene,lelay2015silicene}. Moreover, topological silicon nanowires hosting Majorana fermions could be a materials platform for a quantum computer \cite{frolov2013quantum}. However, like other nanowire candidates, even proximitized ones based on heavier constituents with larger spin-orbit coupling, until now, no conclusive experimental measurements guarantee incontrovertibly the existence of topologically protected Majorana zero modes (MZMs) for the possible realization of qubits \cite{rancic2013exactly,Flensberg2021}. 

\begin{figure}[t]
\centerline{\includegraphics[clip,width=0.50\textwidth,angle=0.]{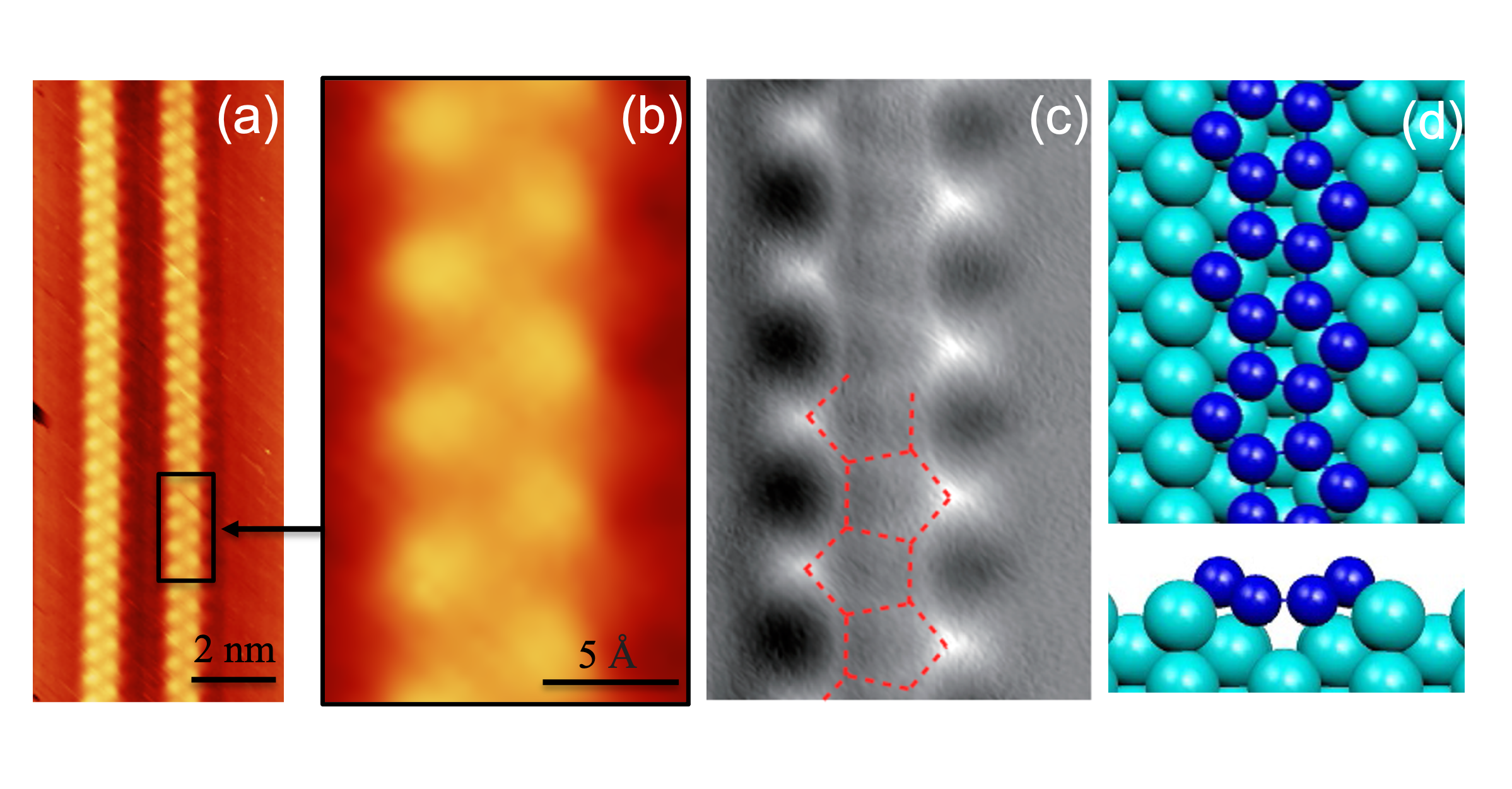}}
    \caption{(Color online) p-SiNR on Ag(110) surface. (a) and (b) Experimental STM images (uncorrected drift), (c) High-resolution nc-AFM image. (d) Top and cross view of the arrangement of the Si pentagonal building blocks. (a) and (b) Courtesy Eric Salomon, (c) Reprinted with permission from \cite{sheng2018pentagonal}. Copyright 2023 American Chemical Society. (d) From Cerda et al.\cite{cerda2016unveiling}.}
    \label{fig:exp}
\end{figure}

Since the appearance of the generic Kitaev model \cite{Kitaev2001}, several platforms were proposed to realize it, both from theoretical \cite{Oreg2010, Aguado17, Schuray2017, Prada2018, Ricco2018, ZhangNatCommun2019, prada2020andreev}, and experimental points of view \cite{Oreg2012, Mourik12, Perge14, Krogstrup2015, JeonScience2017, Clarke2017, Gul2018}. A helpful review of the experimental state-of-the-art on this subject can be found in Refs.~\cite{ LutchynReviewMat2018,Flensberg2021, jack2021}. This model considers \textit{p}-wave superconductor pairing between electrons in different sites of a one-dimensional chain (Kitaev chain) and predicts the existence of unpaired MZMs at opposite ends of a finite Kitaev chain. However, until now, there are no conclusive experimental measurements that guarantee without doubt the existence of topologically protected MZMs~\cite{Pan2020, PanGenericQuantized2020,pan2021quantized,Kim2018, jack2021}. The experimental detection of MZMs remains an elusive problem, and they were not really observed until now. Per se, this situation justifies the search for new platforms.

One possible alternative platform is the one-dimensional honeycomb nanoribbons (HNRs) that have been receiving growing attention in the literature\cite{Jelena_bilayer_graphene_2012, Jelena_nanoribbons_2013, Maiellaro2018, Bento2022}. Nevertheless, the mono-elemental 2D graphene-like materials coined Xenes, where X represents elements from group IIIA to group VIA of the periodic table, could constitute possible candidates to build HNRs with the ability to harbor MZMs at their ends~\cite{Dutreix2014, Ma2017, Aidi20, Grazianetti2021}. Penta-Silicene (X=Si) is an up-and-coming candidate in this family for obtaining a HNR geometry that can host MZMs ~\cite{cerda2016unveiling, sheng2018pentagonal, yue2022}.

A paradigmatic breakthrough would be the experimental implementation of the generic Kitaev toy model with a silicon platform \cite{Kitaev2001}. In a previous work \cite{Bento2022}, we addressed the problem of Majorana spin discrimination employing a double-spin Kitaev zigzag honeycomb nanoribbons (KzHNR), which mimics two parallel Kitaev chains connected by the hopping $t$ (see figure 1 of \cite{Bento2022}). Since such KzHNRs have not been realized in experiments, we look instead in the present paper at the possibility of obtaining MZMs in p-SiNRs, harboring Dirac fermions, which have been epitaxially grown on Ag(110) surfaces~\cite{Depadova2010, DePadova2012,cerda2016unveiling, Iribas2019}. Typically, highly perfect, atom thin, massively aligned single strand p-SiNRs, 0.8 nm in width, and with lengths extending to tens of nanometers were obtained by molecular beam epitaxy upon in situ Si deposition onto Ag(110) surfaces held at room temperature, as shown in Fig. \ref{fig:exp}(a). In scanning tunneling microscopy (STM) and high-resolution nc-AFM images, these p-SiNRs appear as two shifted lines of protrusions along the [110] direction as shown in Fig. \ref{fig:exp}(b-c) and are separated by twice the nearest neighbor Ag-Ag distance, i.e., 0.577 nm. Their hidden internal atomic structure was initially uncovered employing thorough density functional theory (DFT) calculations and simulations of the STM images \cite{cerda2016unveiling}, pointing to an arrangement of pure Si pentagonal building blocks, as displayed in Fig. \ref{fig:exp}(d), which defines the missing pentagonal row (P-MR) model employed in the Supplemental information of reference \cite{cerda2016unveiling} to optimize the angles and the distance between the silicon atoms in the pentagonal arrangement. This unique atomic geometry was later directly visualized by high-resolution non-contact atomic force microscopy (Fig. \ref{fig:exp}(c) from \cite{sheng2018pentagonal}).  We will theoretically demonstrate that these p-SiNRs could constitute a tantalizing disruptive new Kitaev platform.

We propose an experimental implementation for discriminating spin-polarized MZMs in p-SiNRs grown on the Ag(110) surface and aligned along the [110] direction. Since silver is not a superconductor, we will proximitize them with lead, a conventional Bardeen–Cooper–Schrieffer (BCS) superconductor with a relatively high critical temperature of T$_c$ = 7.2 K, upon evaporating \textit{in situ} a thin lead film on top through a mask, as already mentioned in \cite{davila2022silicene}. Indeed, Pb can be easily grown on Ag(110) surfaces \cite{Prince2003} and is known to interact only very weakly with the SiNRs, preserving their integrity and their electronic properties \cite{Krawiec2015,Krawiec2019}. Then, detecting and distinguishing the MZMs at the ends of the SiNRs will be done \textit{in situ} at low temperatures with the STM following the methodology of Yazdani and co-workers \cite{Yazdani2021}. 

In this paper, we characterize the TPTs employing the spinless version of the model and the inclusion of the \textit{p}-wave superconducting pairing and the magnetic field reveal the emergence of topologically protected MZMs with the spin discriminated at opposite ends of the p-SiNRs; this result constitutes the main finding of the work. We also calculate the wave function of the MZMs at the ends of the p-SiNR, showing its topological signature.
\begin{figure}[h]
\centerline{\includegraphics[clip,width=0.50\textwidth,angle=0.]{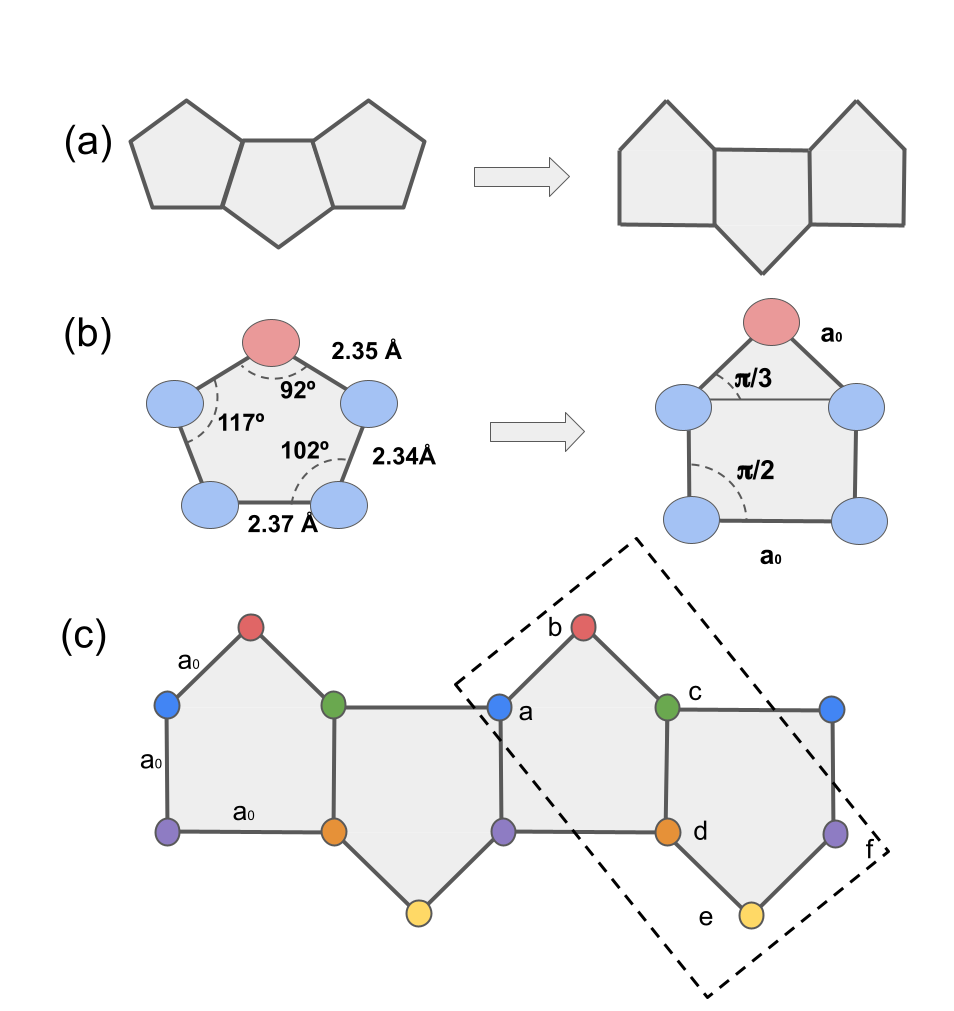}}
\caption{(Color online) (a) Penta-silicene (p-SiNRs) lattice transformation adopted. (b) Penta-silicene angles. (c) Sketch of nonequivalent Si atoms placed at the vertices of the ``square'' pentagonal lattice. In the simulations, we also represent the unit cell employed by the atoms inside the dashed rectangle.}
    \label{fig:LatticeTransformation}
\end{figure}

\section{The Model}

\subsection{Lattice transformations}

In Fig.~\ref{fig:LatticeTransformation}(a), to reduce the geometry complexity of the p-SiNR and facilitate the tight-binding calculations, we redefine its structure using square-shaped pentagons. In the geometry of the pentagons that constitute the p-SiNRs of Fig.~\ref{fig:LatticeTransformation}(b), four silicon atoms are located on the missing pentagonal row, and only one exhibits a buckling structure (pink atoms). We neglect the buckling structure of these atoms and employ a planar configuration composed of square-shaped pentagons. As the distance between the silicon atoms that constitute the pentagons are close, we consider them equal to $a_{0}$ and identify it as the lattice parameter of the p-SiNR. We also define the nearest neighbor hopping as equal to $t$, which is considered the energy unit in all the calculations. $L\equiv 2Na_{0}$, is the length of the p-SiNR, with $a_{0}$ being the distance between atoms and $N$ is the number of sites of the corresponding Kitaev chain (top or bottom), employed in the calculation, as indicated in Fig.~\ref{fig:LatticeTransformation}(c), that exhibits the shape of the p-SiNR and the unit cell composed of six atoms inside the dashed rectangle employed in the calculations. We expect these simplifications will not change the results once we keep the lattice. 
\begin{figure}[h]
\centerline{\includegraphics[clip,width=0.50\textwidth,angle=0.]{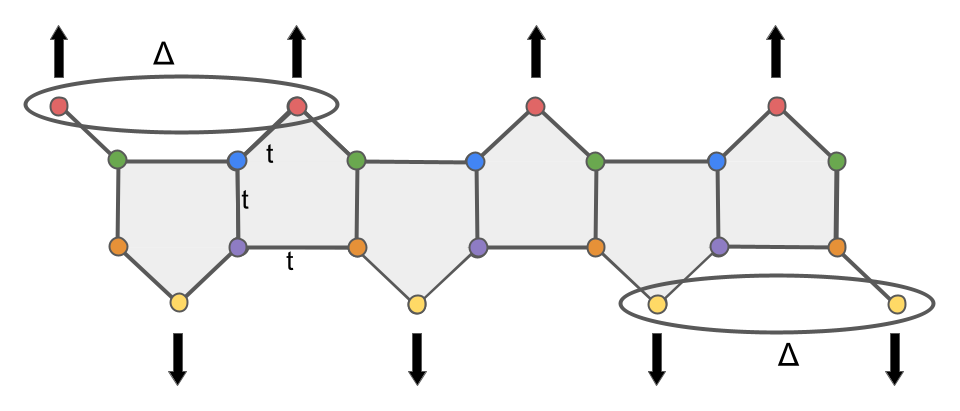}}
\caption{(Color online)  Sketch of the p-SiNRs: The penta-silicene system can be viewed as composed of two Kitaev chains: One on the top and the other on the bottom, hybridized via hopping $t$. The ellipses represent the superconducting \textit{p}-wave pairing between the pink (above) and yellow (below) silicon atoms (in the real material, these atoms correspond to the buckled one). The arrows only indicate the spin polarization needed to define a Kitaev chain.}
    \label{DoubleKitaevchain}
\end{figure}

\subsection{Effective Hamiltonian - spinless case}

The total Hamiltonian, which describes the spinless p-SiNR of Fig. \ref{DoubleKitaevchain}  is given by
\begin{equation}
H =  H_{t} + H_{\Delta} , 
\label{eq:totalHamiltonian}
\end{equation}
with
\begin{equation}
    \begin{split}
        &H_t=-
         \sum_{i=1} ^{N}  \mu \left( a^{\dagger}_{i, +}a_{i, +} - a^{\dagger}_{i, -}a_{i, -}  +b^{\dagger}_{i, +}b_{i, +}-b^{\dagger}_{i, -}b_{i, -} +\right.\\
        &   c^{\dagger}_{i, +}c_{i, +} -c^{\dagger}_{i, -}c_{i, -} + d^{\dagger}_{i, +}d_{i, +}-d^{\dagger}_{i, -}d_{i, -} +  \\
        & \left . e^{\dagger}_{i, +}e_{i, +}-e^{\dagger}_{i, -}e_{i, -} +f^{\dagger}_{i, +} f_{i, +} -f^{\dagger}_{i, -}f_{i, -}  \right ) -\\
        &\sum_{i=1} ^{N} t \left(
        a^{\dagger}_{i, +} b_{i, +} - b_{i, -} a_{i, -}^{\dagger}  +\right.\\
        &b^{\dagger}_{i, +} c_{i, +}- c_{i, -} b_{i, -}^{\dagger} +  
        c^{\dagger}_{i, +} d_{i, +}- d_{i, -} c_{i, -}^{\dagger} +\\
        &\left. d^{\dagger}_{i, +} e_{i, +}- e_{i, -} d_{i, -}^{\dagger} +
        e^{\dagger}_{i, +} f_{i, +}- f_{i, -} e_{i, -}^{\dagger} \right) -\\
        & \sum_{i=1} ^{N-1} t\left(a^{\dagger}_{i+1, +} f_{i, +}- f_{i, -} a_{i+1, -}^{\dagger} +
        a^{\dagger}_{i+1, +} c_{i, +}- \right. \\
       & \left. c_{i, -} a_{i+1, -}^{\dagger} +d^{\dagger}_{i+1, +} f_{i, +}- f_{i, -} d_{i+1,-}^{\dagger}\right) +  \text{H.c.}  ,
    \end{split}
    \label{hamiltonian_real_space_h}
\end{equation}
where $\mu$ is the chemical potential, the index $(-)$ and $(+)$ differentiate the creation and annihilation operators for electrons and holes, respectively, and H.c. is the Hermitian conjugate. The system Hamiltonian of Eq.~(\ref{hamiltonian_real_space_h}) was built according to the unit cell of nonequivalent Si atoms (a,b,c,d,e,f) shown in Fig.~\ref{fig:LatticeTransformation}(c).

The p-SiNRs are grown on Ag(110) surfaces in the setup proposed here. However, silver is not a superconductor, and to generate a \textit{p}-wave pairing $\Delta$ on the pink and yellow  atoms of Fig. \ref{DoubleKitaevchain}, we evaporate in situ a thin lead film over the Ag(110) surface in such a way that the buckled silicon atoms enter in contact with the lead atoms. Under the presence of a strong RSOC arising from the Pb atoms and an applied magnetic field, the \textit{s}-wave Cooper pairs of the Pb film can enter into the p-SiNH region via proximity effect (Andreev reflections)~\cite{Aguado17}, giving rise to a \textit{p}-wave-induced pairing in the double p-HNRs structure. By following the same procedure done in our previous work~\cite{Bento2022} and based on the Kitaev model~\cite{Kitaev2001}, we introduce a spinless  \textit{p}-wave superconducting pairing $\Delta$ between the ``external'' pink and yellow atoms of the same type as shown in Fig.~\ref{DoubleKitaevchain}. The Hamiltonian, which describes such a pairing, reads
\begin{equation}
    \begin{split}
        &H_{\Delta}=
         \sum_{i=1} ^{N-1}  \Delta
        \left(b^{\dagger}_{i,+} b^{\dagger}_{i+1,-} -
        b^{\dagger}_{i+1,-} b^{\dagger}_{i,+}  +\right.\\
        &\left.e^{\dagger}_{i,+} e^{\dagger}_{i+1,-} -
        e^{\dagger}_{i+1,-} e^{\dagger}_{i,+} \right)+ \text{H.c.}  .
    \end{split}
    \label{hamiltonian sup}
\end{equation}

\subsection{Effective Hamiltonian - spinful case}

In order to properly account for the spin degree of freedom in the superconducting p-SiNRs, we follow our previous work~\cite{Bento2022}. We introduce a Zeeman effect due to the application of an external magnetic field perpendicular to the p-SiNRs plane. The Hamiltonian, which accounts for the Zeeman effect, reads:
\begin{equation}
    \begin{split}
        H_z& =  \sum_{i=1, \sigma}^{N}   Z \ sgn(\sigma)
        \left ( a_ {i,\sigma} ^{\dagger} a_{i,\sigma} + 
        b_{i,\sigma} ^{\dagger} b_{i,\sigma} +\right .  \\ 
        & \left . c_{i,\sigma} ^{\dagger} c_{i,\sigma} +
        d_{i,\sigma} ^{\dagger} d_{i,\sigma} + 
        e_{i,\sigma} ^{\dagger} e_{i,\sigma} + 
        f_{i,\sigma} ^{\dagger} f_{i,\sigma} \right ) + \text{H.c.} , \label{eq:H_Z} 
    \end{split}
\end{equation}
wherein $Z \ $ is the effective strength of the external Zeeman field and $\sigma=\uparrow,\downarrow$ is the spin index for each operator. 

The extrinsic RSOC induced in the p-SiNRs can be modulated by the action of an external electric field $\vec{E}$ applied perpendicularly to the nanoribbon plane \cite{Min2006,zarea2009,Ezawa2012,Jiao2020}. Its corresponding general Hamiltonian reads

\begin{equation}
   H_R= \sum_{i,j=1,\sigma}^{N} i c_{i, \sigma} ^{\dagger} (\vec{u}_{i,j} . \vec{\gamma}) c_{j, (\Bar{\sigma})}  + \text{H.c.},\label{eq:H_R} 
\end{equation}
where $\vec{u}_{i,j} = -\frac{R \ }{a_0} \hat{k} \times \vec{\delta}_{i,j}$, with $R \ $ being the extrinsic RSOC parameter, $\vec{\delta}_{i,j}$ is the vector that connects the adjacent lattice sites $i$ and $j$, and $\vec{\gamma}$ the Pauli matrices. The index $\Bar{\sigma}$ indicates the opposite spin direction of $\sigma$. The Eq. \ref{eq:H_R} turns into
\begin{equation}
    \begin{split}
        &H_R = \sum_{i=1,\sigma}^{N}    \left( 
         \gamma_{1}( a^{\dagger}_{i, \sigma}  b_{i+1/2, \Bar{\sigma}} ) +   \gamma_{2} ( b^{\dagger}_{i+1/2, \sigma}   a_{i, \Bar{\sigma}}) +  \right. \\
         &(a^{\dagger}_{i, \sigma} c_{i-1, \Bar{\sigma}} ) -   (c^{\dagger}_{i-1, \sigma}      a_{i, \Bar{\sigma}}) +   
         (-i) ( a^{\dagger}_{i, \sigma}  f_{i, \Bar{\sigma}} ) +  \\
         &(i) (f^{\dagger}_{i, \sigma}  a_{i, \Bar{\sigma}}) +   \gamma_{3} (b^{\dagger}_{i+1/2, \sigma}   c_{i+1, \Bar{\sigma}}) + \gamma_{4} (c^{\dagger}_{i+1, \sigma}  b_{i+1/2, \Bar{\sigma}}) +\\    
         &(-i)(c^{\dagger}_{i+1, \sigma}   d_{i+1, \Bar{\sigma}}) + (i)(d^{\dagger}_{i+1, \sigma}  c_{i+1, \Bar{\sigma}})    -(d^{\dagger}_{i+1, \sigma}      f_{i, \Bar{\sigma}}) + \\  
         & (f^{\dagger}_{i, \sigma}      d_{i+1, \Bar{\sigma}}) + \gamma_{3} (d^{\dagger}_{i+1, \sigma}e_{i+3/2, \Bar{\sigma}}) +  \gamma_{4} (e^{\dagger}_{i+3/2,\sigma}  d_{i+1, \Bar{\sigma}})+\\
         & \left.  \gamma_{1}  (e^{\dagger}_{i+3/2,\sigma}f_{i+2, \Bar{\sigma} }) +  \gamma_{2} (f^{\dagger}_{i+2,\sigma} e_{i+3/2, \Bar{\sigma}}) \right) + \text{H.c.}, 
    \end{split}
\end{equation}
where $ \gamma_{1} = \frac{-1}{2} \  + \frac{i\sqrt{3}}{2}$ , $\gamma_{2} =\frac{1}{2} \  - \frac{i\sqrt{3}}{2}$ , $ \gamma_{3} =\frac{-1}{2} \  - \frac{i\sqrt{3}}{2}$ and $\gamma_{4} =\frac{1}{2} \  + \frac{i\sqrt{3}}{2}$.

Notice that from Eqs.~(\ref{eq:H_Z}) and~(\ref{eq:H_R}), we are assuming the external Zeeman field $Z$ perfectly perpendicular to the RSOC, i.e, $Z\equiv Z_{\perp}\neq 0$ and $Z_{\parallel} = 0$. In Rashba nanowires setups, this condition is responsible for the vanishing of the induced superconducting gap at zero momentum (inner gap) and the opening of a constant gap at finite momentum (outer gap), which characterizes the topological phase transition and the concomitant emergence of MZMs protected by the outer gap~\cite{Aguado17}.

However, from the experimental perspective, ensuring that the magnetic field is applied only in the perpendicular direction of the RSOC field can be challenging. Then, it is natural to consider also the effects of $Z_{\parallel} \neq 0$. In this situation, we have both components of the Zeeman field, and the critical magnetic field condition for the topological phase transition remains the same. However, the behavior of the outer gap is not constant anymore, which affects the topological protection of the MZMs towards fault-tolerant quantum computing operations. The effect of $Z_{\parallel}$ in the outer gap is not so detrimental if the RSOC is strong.

It is worth noticing that the opposite cases of $Z \equiv Z_{\parallel} \neq 0$ and $Z_{\perp} = 0$ can lead to the vanishing of the outer gap, hence preventing the topological phase and emergence of MZMs. Therefore, since our system is qualitatively described by the similar underlying physics of Rashba nanowires, it is appropriate to experimentally ensure the dominance of the magnetic field component perpendicular to the Rashba field.

Considering also the spin degree of freedom on both $H_t$ and $H_{\Delta}$ [Eqs. (\ref{hamiltonian_real_space_h}) and \ref{hamiltonian sup})], we now can define the total system Hamiltonian as
\begin{equation}
    H_{\text{total}} = H_t + H_Z + H_R + H_{\Delta},\label{eq:Htotalspinfull}
\end{equation}
which can be written in the corresponding Bogolyubov-de Gennes (BdG) form in $k$-space as
$H_{\text{total}}(k)=\Phi^{T}\boldsymbol{H}_{\text{BdG}}(k)\Phi$, with
\begin{equation}
\begin{split}
        & \boldsymbol{H}_{\text{BdG}}(k)   = \\ 
        \displaystyle
        & \displaystyle \left[\begin{matrix} 
        H_{\uparrow, \uparrow}(k)& H_{\uparrow, \downarrow}(k)  & H_{\Delta,\uparrow, \uparrow}(k) & H_{\Delta,\uparrow, \downarrow}(k)\\
         H_{\uparrow, \downarrow}(k)& H_{\downarrow, \downarrow}(k)  & H_{\Delta,\downarrow \uparrow}(k) & H_{\Delta,\downarrow, \downarrow}(k) \\
         H^* _{\Delta,\uparrow,\uparrow}(-k) & H^* _{\Delta,\uparrow,\downarrow}(-k) &H^*_{\uparrow, \uparrow}(-k) & H^*_{\uparrow, \downarrow}(-k) \\
         H^* _{\Delta,\downarrow,\uparrow}(-k) & H^* _{\Delta,\downarrow,\downarrow}(-k) &H^*_{\downarrow, \uparrow}(-k) & H^*_{\downarrow, \downarrow}(-k)
         \end{matrix}\right],\label{eq:HBdG}
       \displaystyle
\end{split}
\end{equation}
where $H_{\sigma,\sigma'}(\pm k) $ and $H_{\Delta,\sigma,\sigma'}(\pm k)$ represent the matrix elements for different spin directions and the matrix elements corresponding to the part of the matrix where superconducting couplings $\Delta$ appear, respectively. The spinor $\Phi$ was constructed with the fermionic operators in the following order:
\begin{equation} 
\begin{split}
    \Phi^{T}=(a_{k,\uparrow},b_{k,\uparrow},c_{k,\uparrow},d_{k,\uparrow},e_{k,\uparrow},f_{k,\uparrow},\\
    a_{k,\downarrow},b_{k,\downarrow},c_{k,\downarrow},d_{k,\downarrow},e_{k,\downarrow},f_{k,\downarrow}, \\ a^{\dagger}_{-k,\uparrow},b^{\dagger}_{-k,\uparrow},c^{\dagger}_{-k,\uparrow},d^{\dagger}_{-k,\uparrow},e^{\dagger}_{-k,\uparrow},f^{\dagger}_{-k,\uparrow},\\
    a^{\dagger}_{-k,\downarrow},b^{\dagger}_{-k,\downarrow},c^{\dagger}_{-k,\downarrow},d^{\dagger}_{-k,\downarrow},e^{\dagger}_{-k,\downarrow},f^{\dagger}_{-k,\downarrow}) .
\end{split}    
\end{equation}

The spin alignment for each situation in the next section is computed numerically. We calculate the mean value of the Pauli matrix in $\hat{z}$ direction $\hat{S}_z$, i.e., $\langle \hat{S}_z \rangle = \langle\Psi|\hat{S}_z|\Psi\rangle$, where $|\Psi\rangle$ are the eigenvectors of the total Hamiltonian given by Eq.~(\ref{eq:Htotalspinfull}).

In hybrid semiconducting-superconducting nanowires, sometimes called Majorana nanowires, the following features strongly suggest the emergence of MZMs  at the nanowire ends~\cite{Aguado17}:
\begin{itemize}
\item[(a)] Closing and subsequent  reopening of the superconducting gap in the bulk relation dispersion as the chemical potential $\mu$ changes, indicating a TPT;
\item[(b)] Emergence of persistent zero-modes for specific system parameter values associated with nonoverlapping wave functions localized at the opposite ends of the nanowire.
\end{itemize}

To obtain the TPTs present in the p-SiNRs, we will consider the infinite case given by the Hamiltonian of Eq.~(\ref{eq:totalHamiltonian}). We calculate the bulk band structure, discussed in detail in the supplemental material (SM). To investigate the existence of MZMs in the p-SiNRs, we will analyze the spinless p-SiNRs with finite size $N=100$ and calculate the energy spectrum as a function of the chemical potential $\mu$ and the probability density function $|\psi|^2$ associated with the zero-energy states which arise on the real axis of the energy spectrum.

Both the energies $E_{n}$ and eigenvectors $\psi_{n}$ per site are obtained by numerically solving the Schrödinger equation $ H \psi_n = E_n \psi_n$ for the Hamiltonian of Eq.~(\ref{eq:totalHamiltonian}). To evaluate the position dependence of the wave functions associated with zero energy states, we numerically calculate the eigenvector $\psi_{n}$ when $E_{n}=0$, which allows obtaining the probability density per lattice site according to
\begin{equation}
    | \psi_n | ^2 = \psi_n \psi_n ^{*}. \label{eq:probability density}
\end{equation}
\begin{figure}[t]

\centerline{\includegraphics[clip,width=0.58\textwidth,angle=0.]{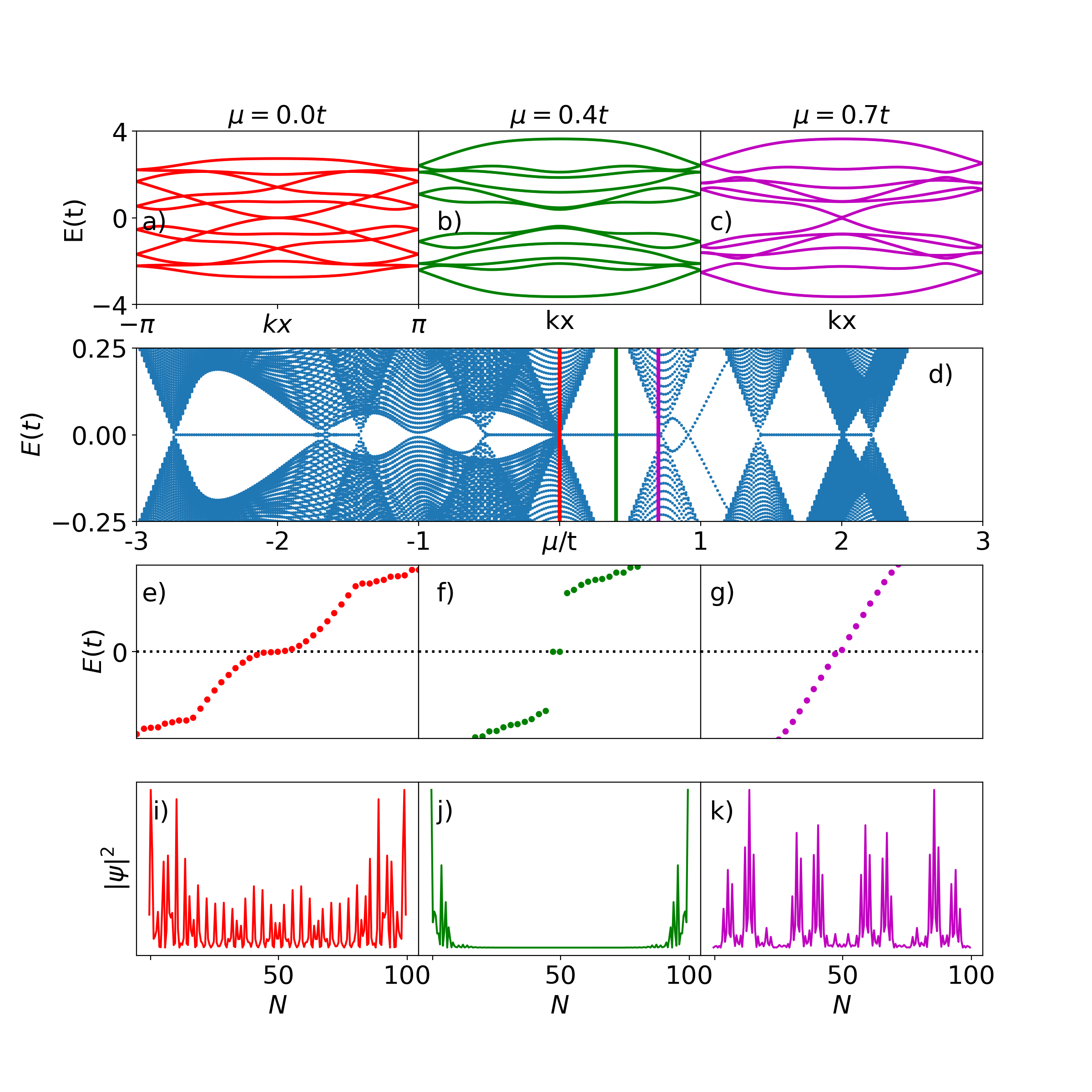}}
	\caption{(Color online) \textbf{Spinless case:} \label{fig:Result1}\textbf{(a)-(c)} Bulk energy dispersion for the spinless p-SiNRs as a $k_{x}$ function. The colors red, green, and magenta used in the panels correspond to $\mu=0.0t$, $0.4t$, and $0.7t$, respectively. \textbf{(d)} Energy spectrum as a function of the chemical potential. \textbf{(e)-(g)} Zero-energy states spectrum. \textbf{(i)-(k)} Probability density per lattice site $|\psi|^2$, associated with zero-energy states on the real axis of the Kitaev, top or bottom chains.}
\end{figure}

\section{Results and discussion}
\subsection{Finite spinless p-SiNRs}

We employed the following parameter set in all the calculations: $\Delta = 0.5 t$, $Z=0.1t$, $R=0.05t$ and $N=100$. The top panels of Fig.~\ref{fig:Result1} show the bulk energy dispersion of the p-SiNRs, in the presence of the superconducting \textit{p}-wave pairing, described by Eqs.~(\ref{eq:totalHamiltonian}-\ref{hamiltonian sup}), along the $k_{x}$ direction, for three representative values of chemical potential $\mu$ [vertical arrows in panel (d)]. Fig. \ref{fig:Result1}(a) depicts the closing of the SC gap at $k_{x}=0$ for 
$\mu = 0.0t$. As the value of $\mu$ enhances, the SC gap opens as shown in panels (b) for 
$\mu = 0.4t$ and closes again at $k_{x}=0$ for $\mu = 0.7t$ as shown in panel (c). This closing and reopening of the SC gap with the tuning of $\mu$ characterize a topological phase transition. The bulk-boundary correspondence principle \cite{Alase2019} ensures the topologically protected MZMs at the ends of the p-SiNRs.

To verify the emergence of MZMs associated with the TPTs depicted in Fig. \ref{fig:Result1}(a)-(c), we plot the p-SiNRs energy spectrum as a function of $\mu$ in Fig.~\ref{fig:Result1}(d). There are no zero-energy modes for the values of $\mu$ where the gap closes (red and magenta vertical lines). However, for values of $\mu$ inside the topological gap, for example, when $\mu=0.4t$ (green vertical line), two zero-energy states appear on the real axis, indicating the presence of MZMs at the opposite ends of the p-SiNRs, topologically protected by the effective \textit{p}-wave SC gap [Fig.~\ref{fig:Result1}(b)]. This finding is similar to what was obtained in our previous work~\cite{Bento2022}, wherein the MZMs emerge at the opposite ends of a finite double zHNR.

Fig.~\ref{fig:Result1}(f) shows isolated zero-energy modes for $\mu=0.4t$, which are associated with a nonoverlapping wave function, well-localized at the ends of the p-SiNRs, as depicted in Fig.~\ref{fig:Result1}(j); which together with the topological phase transition [Fig.~\ref{fig:Result1}(a)-(c)], is a piece of strong evidence that topologically protected MZMs emerge at the opposite ends of the spinless p-SiNRs. In the Supplemental Material, we developed an extensive analysis of the topological and trivial phases of the spinless \textit{p}-wave superconducting p-SiNR, that can be distinguished by the Zak number topological invariant~\cite{ZakPhysRevLett.62.2747(1989)}. However, we cannot afford to do the same study for the spinful case due to the extreme mathematical complexity.

Although there are zero-energy modes for other values of $\mu$ [Fig.~\ref{fig:Result1}(e) and (g)], they are not associated with wave functions well-localized at the ends of the p-SiNR, as can be seen in Figs.~\ref{fig:Result1}(i) and (k), for $\mu=0.0t$ and $\mu=0.7t$, respectively.

We also highlight that we analyze only one  region of all energy spectrum shown in Figs.~\ref{fig:Result1}(d), which presents other ranges of chemical potentials wherein a zero-energy state, associated with the emergence of MZMs, arises. A more detailed study of this energy spectra can be found in the SM. We can also observe that, unlike the system of our previous work~\cite{Bento2022}, the energy spectrum of Figs.~\ref{fig:Result1}(d) is asymmetric at about $\mu = 0$.

\begin{figure}[t]
\centerline{\includegraphics[clip,width=0.54\textwidth,angle=0.]{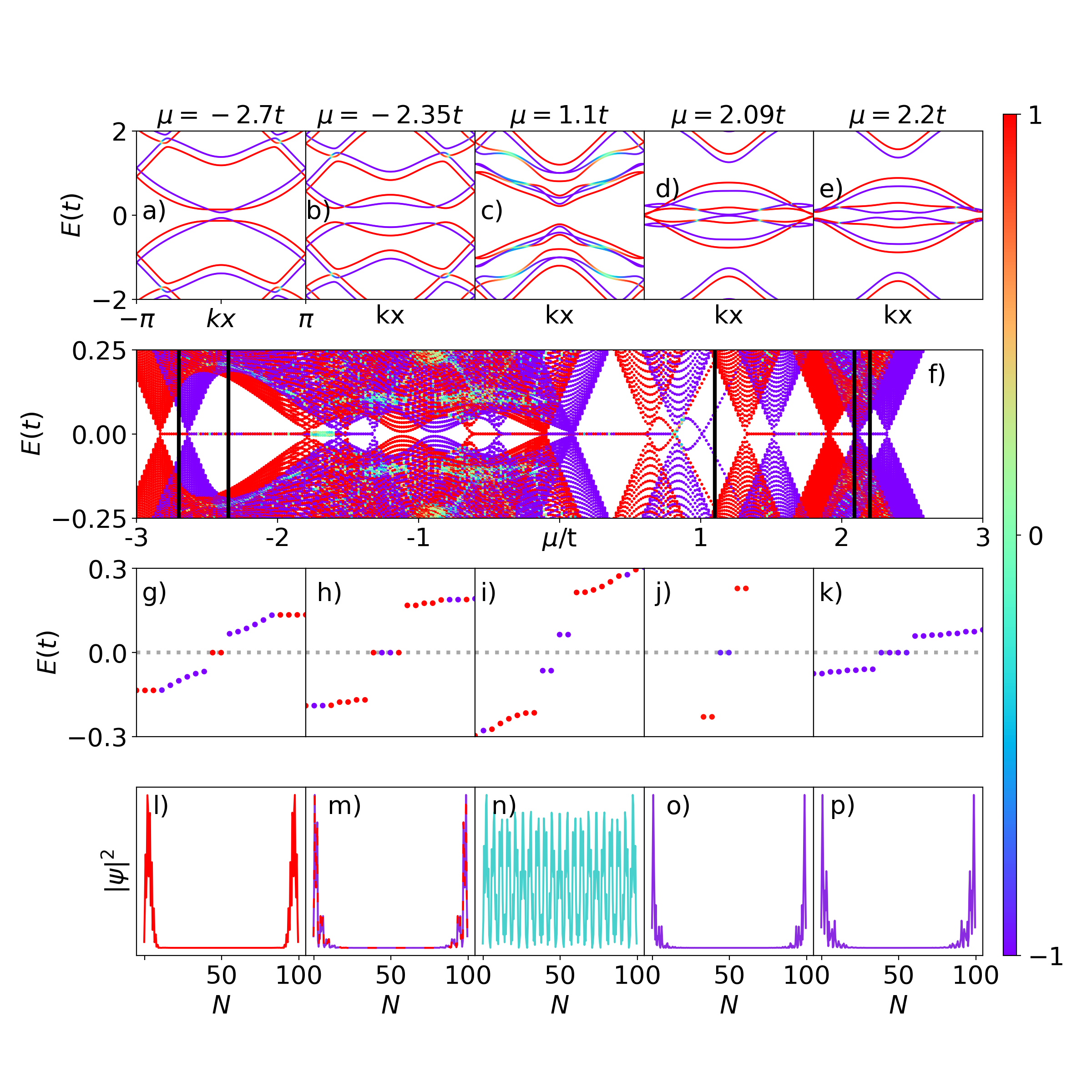}}
\caption{(Color online) \textbf{Spinful case - Magnetic field up:} \label{fig:Result22}\textbf{(a)-(e)} Bulk energy dispersion of the superconducting p-SiNRs for the spinful situation, as a function of $k_{x}$, for $\mu=-2.7t$, $-2.35t$, $1.1t$, $2.09t$ and $2.2t$, respectively. \textbf{(f)} Energy spectrum as a function of the chemical potential. Vertical lines indicate the chosen values of chemical potential shown on top panels. \textbf{(g)-(k)} Zero-energy states spectrum. \textbf{(l)-(q)} Probability density per lattice site $|\psi|^2$, associated with zero-energy states on the real axis of the spin-polarized Kitaev, top or bottom chains.}
\end{figure}

\begin{figure}[t]
\centerline{\includegraphics[clip,width=0.54\textwidth,angle=0.]{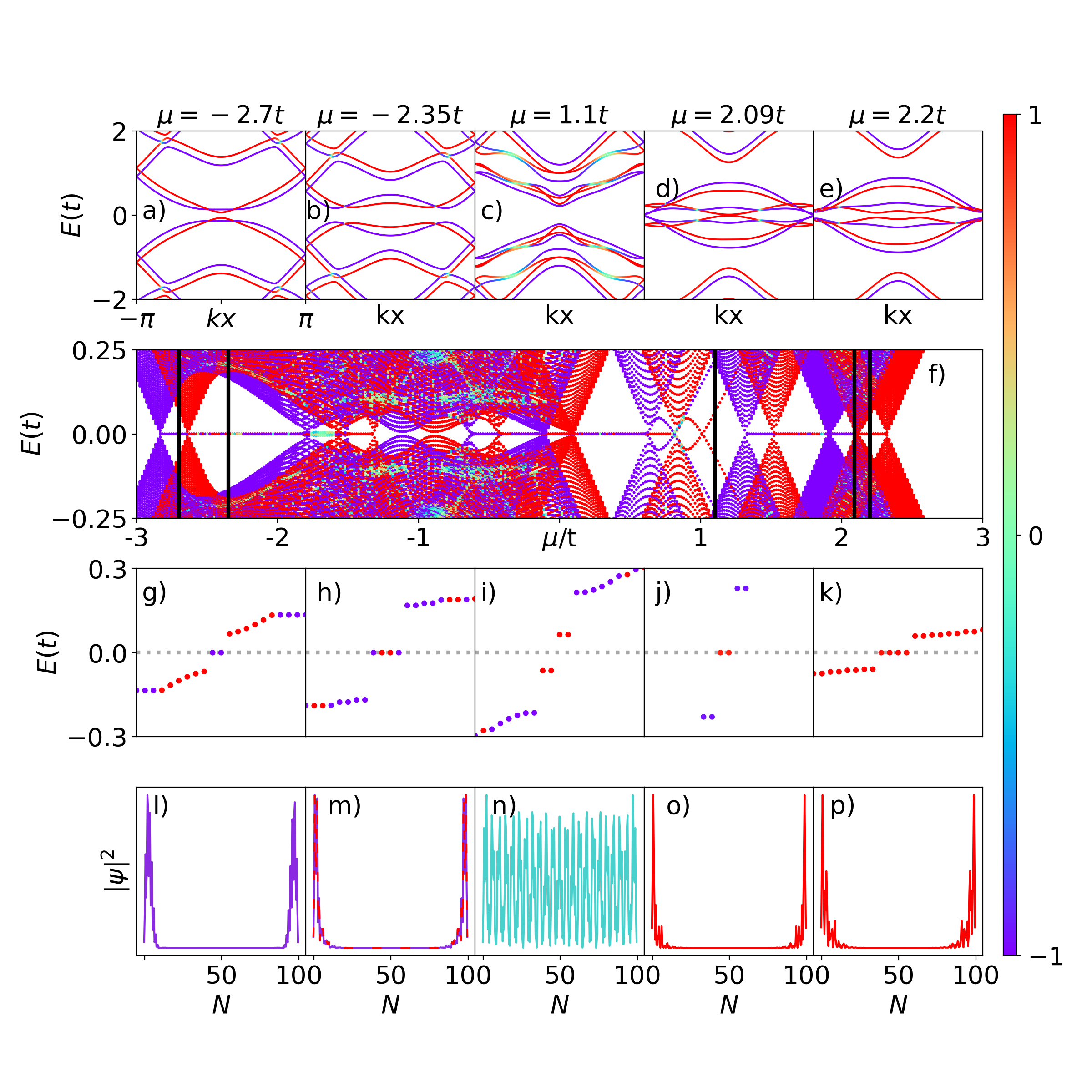}}
\caption{(Color online) \textbf{Spinful case - Magnetic field down:} \label{fig:Result222} The same situation of Fig. \ref{fig:Result22} but with the magnetic field pointing in the opposite direction.}
\end{figure}

\subsection{Finite spinful p-SiNRs}

Now we will analyze how the spinless scenario shown in Fig.~\ref{fig:Result1} is affected by the presence of both Zeeman field [Eq.~(\ref{eq:H_Z})] and extrinsic RSOC [Eq.~(\ref{eq:H_R})] coupling within the spinful description [Eq.~(\ref{eq:Htotalspinfull})]. 

Figs.~\ref{fig:Result22}(a)-(e) exhibit the energy dispersion of the p-SiNRs given by the eigenenergies of BdG Hamiltonian [Eq.~(\ref{eq:HBdG})] as a function of $k_{x}$, for distinct values of the chemical potential $\mu$, indicated by vertical lines in Fig.~\ref{fig:Result22}(f). The spin polarization is indicated by the vertical color bar, in which the red color represents the spin $\uparrow=1$, while the blue color stands for spin $\downarrow=-1$, and the light shades of colors mean the spin is neither up nor down. As $\mu$ is tuned, we can see the opening and closing of the superconducting gap, thus indicating a TPT, as previously verified in the spinless situation [Fig.~\ref{fig:Result1}(a)-(c)]. However, here we can notice that each TPT associated with a specific value of $\mu$ has a preferential spin orientation, except Fig~\ref{fig:Result22}(c), where the system exhibits a conventional band gap.

The spin-polarized TPTs in Figs.~\ref{fig:Result22}(a),(b),(d), and (e) lead to the appearance of spin-polarized zero-modes in Fig.~\ref{fig:Result22}(f), which shows the system energy spectrum as a function of $\mu$. These zero-modes indicate the emergence of spin-polarized MZMs at the ends of the p-SiNRs as $\mu$ is changed, similar to those found in~\cite{Bento2022}. 

The panels (g)-(k) of Fig.~\ref{fig:Result22} depict the corresponding energy levels sorted in ascending order. The different values of $\mu$ used to calculate the MZMs are indicated by vertical black lines in Fig.~\ref{fig:Result22}(f). For $\mu=-2.7t$ [Fig.~\ref{fig:Result22}(g)], there are two zero modes on the real axis of spin up (red points), associated with nonoverlapping wave functions shown in Fig.~\ref{fig:Result22}(l). For $\mu=-2.35t$ [Fig.~\ref{fig:Result22}(h)], there are two energy-states in the spin-up direction and other two with spin-down, associated with degenerate (blue and red) nonoverlapping wave functions shown in Fig.~\ref{fig:Result22}(m). For $\mu=1.1t$ [Fig.~\ref{fig:Result22}(h)], there are four spin-down energy states outside the real axis, there are no MZMs, and the wave functions completelly overlap along the ribbon. For $\mu=2.09t$ [Fig.~\ref{fig:Result22}(j)], there are two zero modes on the real axis of spin up (red points). Finally, for $\mu=2.2t$ [Fig.~\ref{fig:Result22}(k)], there are four MZMs with spin-down energy states on the real axis. This situation happens because, at $\mu=2.09t$, a TPT occurs for spin-up, the gap closes at $k=\pm \pi$, and for $\mu>2.09t$ the gap defines a trivial band insulator for this spin orientation and MZMs with spin-up are not available anymore. These well-localized probability densities describing wave functions centered at the opposite ends of the superconducting p-SiNRs, associated with zero-energy edge states, indicate the emergence of MZMs in the same way previously found for the spinless system.

Fig. \ref{fig:Result222} represents the same situation as Fig. \ref{fig:Result22} but with the magnetic field pointing in the opposite direction. The net effect on the p-SiNRs is to change the MZMs, for all $\mu$ values, in spin up to down and vice versa. Therefore, it is possible to select the spin polarization of the MZMs by changing the chemical potential $\mu$ or the magnetic field orientation.

\begin{figure}[t]
\centerline{\includegraphics[clip,width=0.50\textwidth,angle=0.]{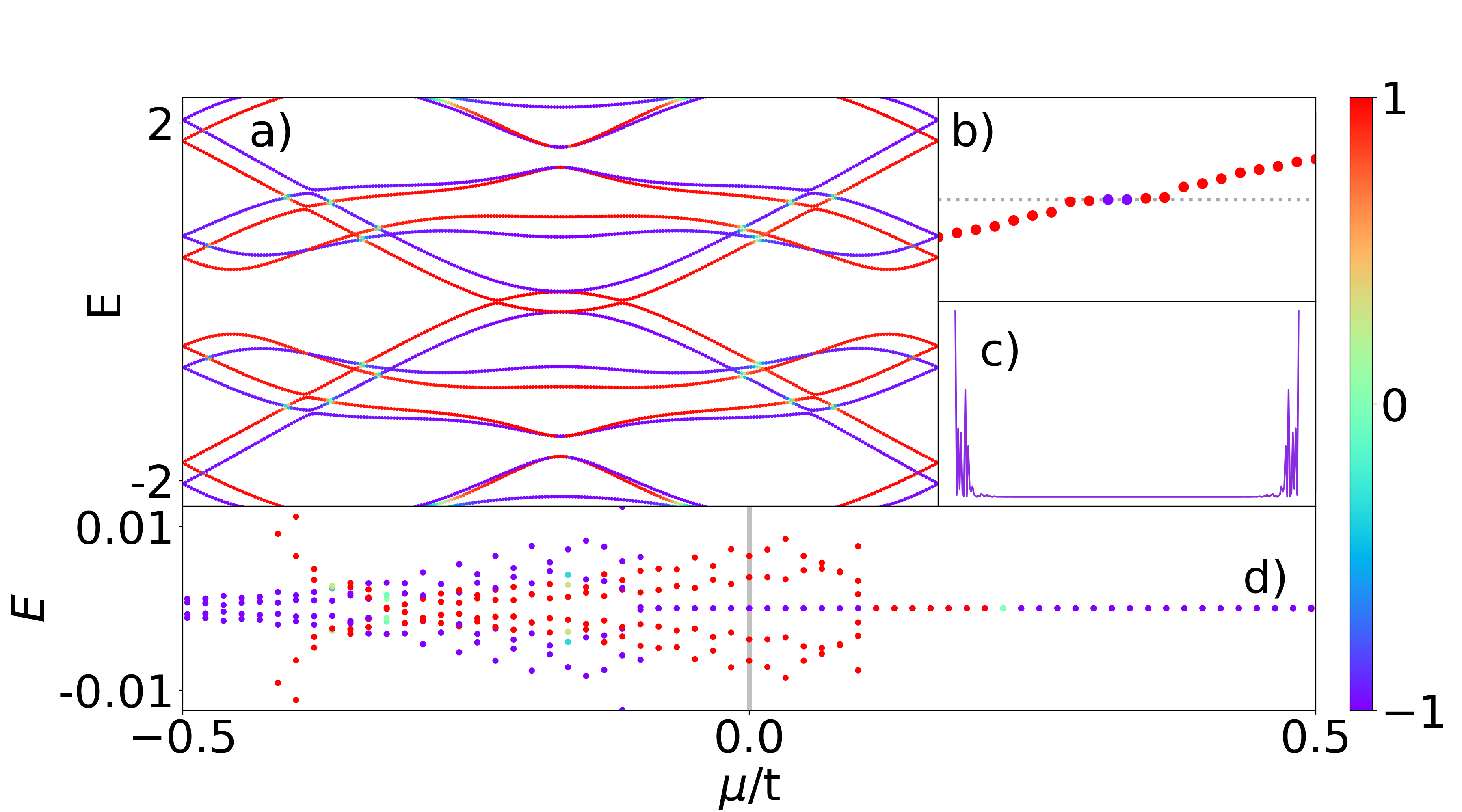}}
\caption{(Color online) \label{fig:mu0} Analysis in detail of the $\mu=0$ case of Fig. \ref{fig:Result22} with the magnetic field pointing in the up direction.}
\end{figure}
\begin{figure}[t]
\centerline{\includegraphics[clip,width=0.48\textwidth,angle=0.]{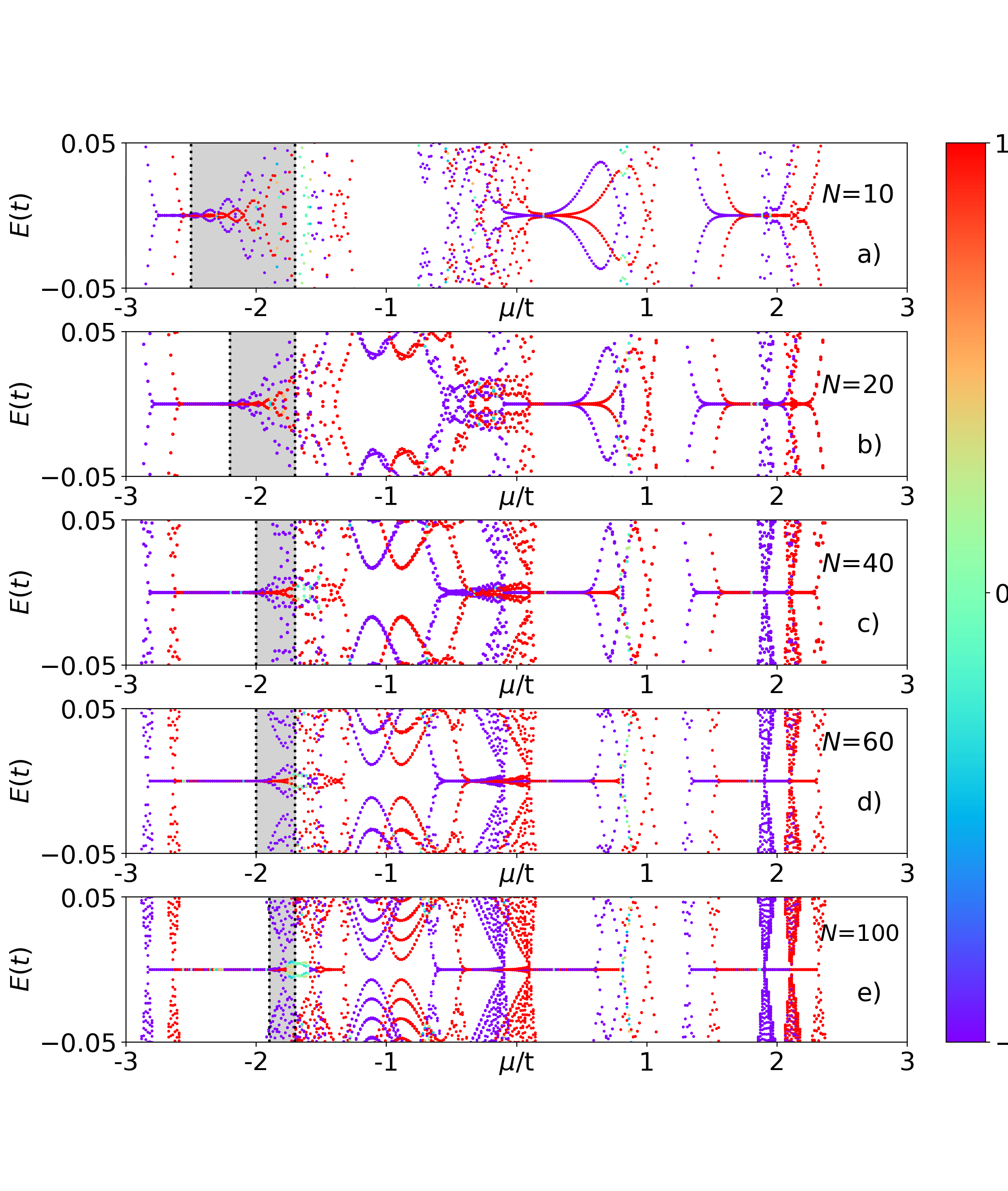}}
\caption{(Color online) \label{fig:Result4} Energy spectrum as a function of the chemical potential $\mu$, for distinct lengths of superconducting p-SiNRs, namely, for $N=10$ \textbf{(a)}, $N=20$ \textbf{(b)}, $N=40$ \textbf{(c)}, $N=60$ \textbf{(d)} and $N=100$ \textbf{(e)}.} 
\end{figure}

In Fig.~\ref{fig:mu0}, we mainly analyze the dispersion relation, energy spectrum, and nature of the zero-modes at $\mu=0$ of Fig. \ref{fig:Result22}, with the magnetic field pointing in the up direction. Fig.~\ref{fig:mu0}(a) depicts $E(k)$ as a function of $k_x$, showing that there is a finite topological superconducting gap only for the spin-down orientation (blue line), while the spin-up (red line) remains gapless. This behavior suggests a spin-polarized TPT at zero chemical potential, implying that only the system's spin-down component is within the topological regime. At the same time, the spin-up belongs to a metallic phase. Fig.~\ref{fig:mu0}(b),(c) represent two MZMs of spin-down with its correspondent nonoverlapped wave function, respectively, and Fig.~\ref{fig:mu0}(d), shows detail at around $\mu=0$ region.

We also investigate how the energy spectrum as a function of $\mu$ is affected by the length of the p-SiNRs. Fig.~\ref{fig:Result4} exhibits the energy spectrum of the superconducting p-SiNRs for increasing values of nanoribbon length $N$. From the smallest system considered [$N=10$, Fig.~\ref{fig:Result4}(a)] to the largest one [$N=100$, Fig.~\ref{fig:Result4}(e)], it can be noticed a decrease of the amplitude of oscillations at around the real axis ($E=0$), and at the same time the definition of the MZMs on the real axis improves as $N$ increases, and for $N=100$ the MZMs are well defined in all the real axis. It should be mentioned that these oscillations around zero energy are expected for short Majorana nanowires due to the overlap between Majorana wavefunctions of opposite ends. Therefore, such oscillations are expected to decrease as the system becomes larger. The same behavior was verified in the work~\cite{Bento2022}.

\section{Conclusions and Perspectives}

This paper demonstrates the emergence of topologically protected MZMs at opposite ends of spinless and spinful p-SiNRs with \textit{p}-wave superconducting pairing. These MZMs exhibit spin discrimination, and their polarization can be controlled by adjusting the nanoribbon chemical potential or the external magnetic field. To implement our findings experimentally, we propose a material engineering of p-SiNRs grown over an Ag(110) surface [cf. Fig.~\ref{fig:exp}(a)], with a thin Pb film deposited on top~\cite{davila2022silicene,Prince2003}. In this device, the proximity effect will enable the penetration of Cooper pairs from the Pb \textit{s}-wave superconductor into the p-SiNRs~\cite{Aguado17}, and in combination with an external magnetic field and the extrinsic RSOC modulated by the action of an external electric field $\vec{E}$ applied perpendicularly to the nanoribbon plane \cite{Min2006,zarea2009,Ezawa2012,Jiao2020}, it will induce \textit{p}-wave pairing in the buckled atoms of the double p-SiNRs structure [cf. Fig.~\ref{fig:exp}(d)].

We should highlight the potential applications driven by the spin-polarized MZMs presented in this work, notably demonstrated in the results of Fig.~\ref{fig:mu0}, with the down spin component associated with MZMs, while the up component displays metallic features, resulting in a half-metallic behavior for the system~\cite{Tsai2013, Jiang_2019}. This property could be harnessed to design a single Majorana transistor (SMT) built from a quantum dot (QD) sandwiched by finite p-SiNR leads ~\cite{Ensslin2010,lelay2015silicene,Niu2023}. This setup resembles the conventional single electron transistor (SET)~\cite{Goldhaber1998}. The SMT can be a valuable tool for discerning between MZMs and trivial Andreev bound states~\cite{prada2020andreev,Ricco2018}. Particularly, the leakage of MZMs through the QD~\cite{VernekLeakagePhysRevB.89.165314(2015)}, along with both local and crossed Andreev reflections induced by a specific spin orientation within the p-SiNR-QD-p-SiNR SMT structure, is expected to generate distinct electronic transport signatures, enabling the identification of MZMs.

In addition to the spin-polarization of MZMs, our proposal also features the emergence of four MZMs at the ends of the p-SiNR, as illustrated in Figs.~\ref{fig:Result22}(h,k) and \ref{fig:Result222}(h,k). Two MZMs are located at opposite ends of the top chain, while another two are at the bottom. Depending on the chemical potential and applied magnetic field orientation, these MZMs can exhibit either the same or opposite spin orientations. Having four MZMs, at least, is crucial for implementing quantum computing operations between two qubits, as it requires the presence of two fermionic sites, i.e., four MZMs \cite{KarzigPRB2017,Steiner2020}. Therefore, our proposal is a promising candidate for realizing hybrid quantum computing operations \cite{LeijnsePRL2011,Flensberg2012} between conventional qubits and spin-polarized Majorana-based qubits and paves the way for defining quantum computing operations using Majorana spintronics \cite{Sarma2016}.

\section{Appendix}
\label{appendix}

\subsection{Topological Classification and Zak phase topological invariant}

The classification of the topological phases of matter is provided by the analysis of fundamental symmetries for a given Hamiltonian in the discrete reciprocal space~\cite{BernevigBook,Chiu2016,StanescuBook}, namely time-reversal ($\mathcal{TR}$), particle-hole ($\mathcal{PH}$) or charge conjugation and chiral symmetries ($\mathcal{K}$). 

For the particular case of the spinless penta-silicene nanoribbons (p-SiNRs) with \textit{p}-wave superconducting pairing at their edges [Eq.~(\textcolor{blue}{1-3}) of main text], it is verified that both $\mathcal{TR}$ and $\mathcal{PH}$ symmetries are preserved, once
\begin{equation}
\mathcal{T}h(k) \mathcal{T}^{-1} = h(-k)\label{time_reversal} 
\end{equation}
and
\begin{equation}
\mathcal{C}h(k)\mathcal{C}^{-1}=-h(-k),
\label{charge_conjugation}  
\end{equation}
where $\mathcal{T}$ and $\mathcal{C}$ are the time-reversal and charge conjugation operators, respectively, and $h(k)$ is a matrix coming from the Hamiltonian of Eq.~(\textcolor{blue}{1-3}) in the manuscript, rewritten in the Bogoliubov-de Gennes (BdG) representation, i.e.,
\begin{equation}
\mathcal{H}(k)= \frac{1}{2}\sum_{k} \Psi^{\dagger}_{k} h(k) \Psi_{k},
\label{eq:HBdG_Spinless}   
\end{equation}
with
\begin{equation}
 \Psi_{k} \equiv(a_{k}, a^{\dagger}_{-k},  b_{k},b^{\dagger}_{-k}, c_{k}, c^{\dagger}_{-k}, d_{k},  d^{\dagger}_{-k}, e_{k},  e^{\dagger}_{-k} ,f_{k},  \\f^{\dagger}_{-k})^{T}\label{eq:BdG_Spinor}    
\end{equation}
being the spinor, which accounts the assumption of $\mathcal{PH}$ symmetry. 

The fulfilment of both $\mathcal{TR}$ and $\mathcal{PH}$ symmetries directly implies that the $\mathcal{K}$ symmetry is also preserved~\cite{StanescuBook}, meaning that
\begin{equation}
\mathcal{K}h(k)\mathcal{K}^{-1} = -h(k), \label{Chiral} 
\end{equation}
where $\mathcal{K}= \mathcal{T} \cdot \mathcal{C}$ corresponds to the chiral operator. Moreover, from the relations expressed in Eqs. (\ref{time_reversal}), (\ref{charge_conjugation}) and (\ref{Chiral}), we obtain $\mathcal{T}^{2}=1$, $\mathcal{C}^{2}=1$ and $\mathcal{K}^{2}=1$, meaning that the BdG Hamiltonian [Eq.~(\ref{eq:HBdG_Spinless})] of the spinless superconducting p-SiNR [Eq.~(\textcolor{blue}{1-3}), main text] is a representative of the BDI symmetry class~\cite{StanescuBook}, the same class of the well-known Kitaev chain~\cite{Kitaev2001}.

It is worth mentioning that spinless superconducting p-SiNR is a simplification considering an ``intrinsic'' magnetic field. The presence of this field is crucial for inducing the formation of \textit{p}-wave superconducting pairing along the nanoribbon edges. However, in practical experimental setups, the source of the spin-polarization is an external magnetic field that naturally breaks the $\mathcal{TR}$ symmetry and hence, $\mathcal{K}$ symmetry. From this argument, the ``artificial'' $\mathcal{TR}$ symmetry of the spinless model can be neglected. Thus the BdG Hamiltonian of Eq.~(\ref{eq:HBdG_Spinless}) belongs to the D symmetry class~\cite{StanescuBook,BernevigBook}. Therefore, the p-SiNR in presence of an applied magnetic field is a $\mathbb{Z}_{2}$ superconductor in one-dimension~\cite{Aguado17,StanescuBook}, once the proposed double-spin Kitaev zigzag nanoribbon configuration can be regarded as two interconnected Kitaev chains with a hopping term (cf. discussion in the main text).

From the previously discussed perspective, the topological and trivial phases of the spinless \textit{p}-wave superconducting p-SiNR, as described by the BdG Hamiltonian of Eq.~(\ref{eq:HBdG_Spinless}), can be distinguished by the Zak number topological invariant~\cite{ZakPhysRevLett.62.2747(1989)}
\begin{equation}
\varphi_{\text{Zak}}= - \int_{-\pi}^{\pi} \frac{dk}{2 \pi i} \partial _k \ln{[\text{Det}(A(k))]}.\label{eq:Zak} 
\end{equation}
A nonzero quantized Zak phase $\varphi_{\text{Zak}}$ is associated with the emergence of topologically protected edge states, which is an outcome of the conventional bulk-boundary correspondence~\cite{StanescuBook,Chiu2016}. Specifically, the integer values of $\varphi_{\text{Zak}}$ topological invariant correspond to the number of topologically protected edge modes present in the system and characterize its topological phase transitions (TPTs).

To compute the Zak number through Eq.~(\ref{eq:Zak}), it is necessary to obtain a chiral matrix $\mathcal{A}(k)$ associated with $h(k)$, which is performed through the computation of a unitary transformation outlined below:
\begin{equation}
 \tilde{h}(k) = \mathcal{U}^{\dagger}h(k)\mathcal{U} = \begin{bmatrix}0 & A(k)\\
A^{*}(k) & 0
\end{bmatrix},  \label{Eq:ChiralForm}  
\end{equation}
bringing $h(k)$ to its chiral form, where
\begin{equation}
\begin{split}
   & A(k)= \\
&\displaystyle \left[\begin{matrix}2 \mu & 2 t e^{\frac{i k}{2}} & 2 t e^{- i k} & 0 & 0 & 2 t\\2 t e^{- \frac{i k}{2}} & 2 \Phi_{k} + 2 \mu & 2 t e^{\frac{i k}{2}} & 0 & 0 & 0\\2 t e^{i k} & 2 t e^{- \frac{i k}{2}} & 2 \mu & 2 t & 0 & 0\\0 & 0 & 2 t & 2 \mu & 2 t e^{\frac{i k}{2}} & - 2 t e^{- i k}\\0 & 0 & 0 & 2 t e^{- \frac{i k}{2}} & 2 \Phi_{k} + 2 \mu & 2 t e^{\frac{i k}{2}}\\2 t & 0 & 0 & - 2 t e^{i k} & 2 t e^{- \frac{i k}{2}} & 2 \mu\end{matrix}\right],\label{eq:A(k)}
\end{split}
\end{equation}
is the chiral matrix, with $\Phi_{k} = i \Delta \sin{\left(2 k \right)}$.

\begin{figure}[h]
	\centerline{\includegraphics[width=4.0in,keepaspectratio]{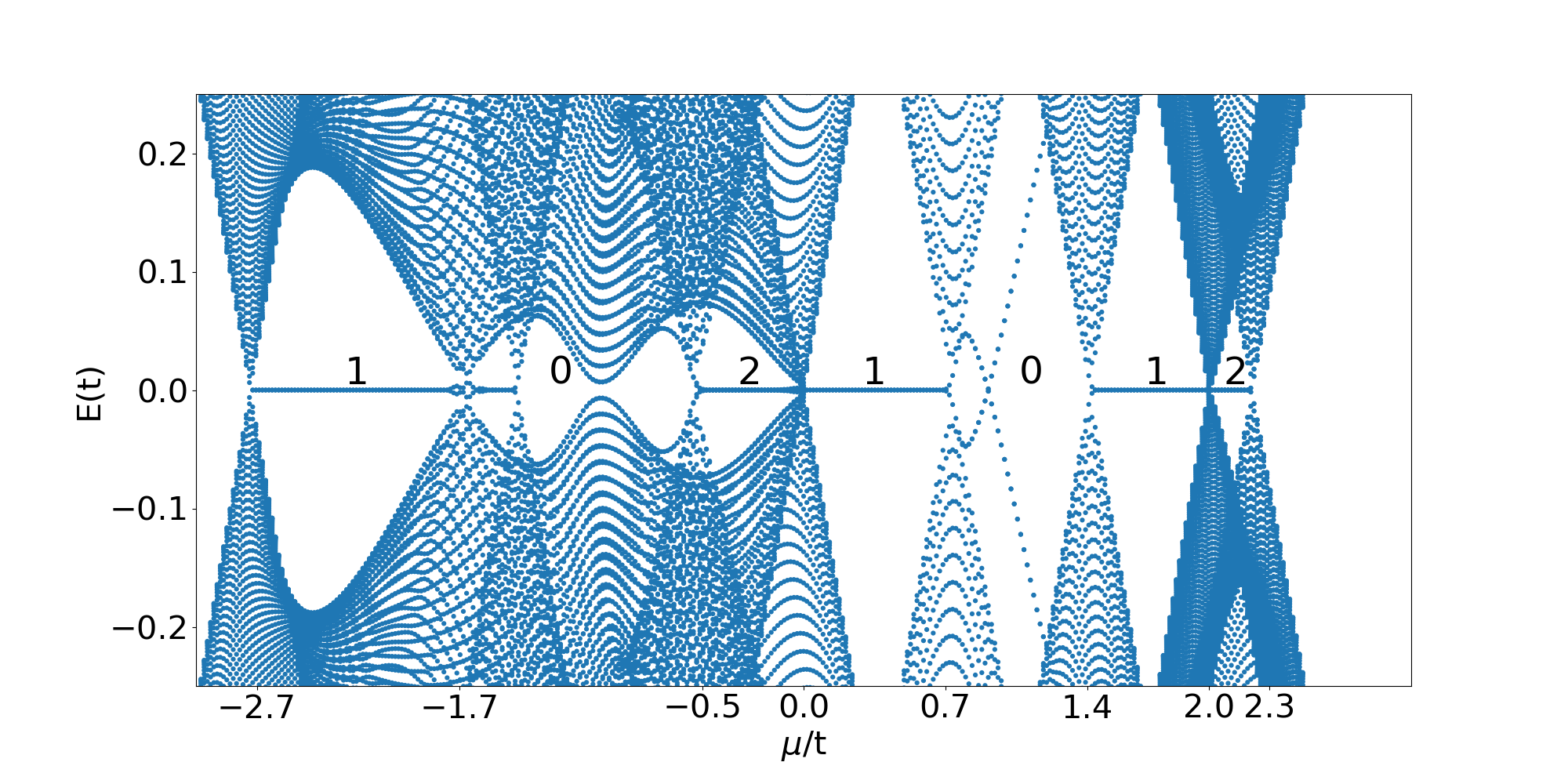}}\caption{Energy dispersion of the bulk system as a function of the chemical potential $\mu$, for the spinless p-SiNR with \textit{p}-wave superconducting pairing between the atoms localized at the edges, cf. Eq.~(1) of the main text. The Zak phase $\varphi_{\text{Zak}}$, represented by the values $0$, $1$, and $2$, corresponds to the number of MZMs present at the edges of either one or both chains comprising the p-SiNR.  
	\label{fig:zaknumber}}
\end{figure}

By considering Eq.~(\ref{eq:A(k)}) and Eq.~(\ref{eq:Zak}) and employing numerical integration, it becomes feasible to compute the Zak number for several values of chemical potential $\mu$. The manipulation of $\mu$ triggers the closing and subsequent reopening of the superconducting gap, a phenomenon closely related to the TPTs, as discussed in the main text.

In this context, Fig.~\ref{fig:zaknumber} illustrates the Zak number across distinct regions in the bulk energy dispersion of the spinless p-wave superconducting p-SiNR. Notably, a Zak phase of zero corresponds to regions where zero modes are absent, indicating that the system resides within the topologically trivial phase. Conversely, for $\varphi_{\text{Zak}}\neq 0$, zero-energy modes emerge, indicating the presence of topologically protected Majorana zero modes (MZMs) at the edges of either one ($\varphi_{\text{Zak}}= 1$) or both top/bottom chains ($\varphi_{\text{Zak}} = 2$) of the p-SiNR~\cite{Bento2022}.

\section*{Data availability}

The data that support the findings of this study are available from the corresponding author upon reasonable request.

\section*{Acknowledgements}

M~S~F, M~A~C, and A~C~S acknowledge financial support from the National Council for Scientific and Technological Development (CNPq) grant numbers 311980/2021-0, 305810/2020-0, 308695/2021-6, respectively. M~S~F acknowledges the Foundation for Support of Research in the State of Rio de Janeiro (FAPERJ) processes number 210 355/2018 and 211.605/2021. M~A~C acknowledges financial support to the Foundation for Support of Research in the State of Rio de Janeiro (FAPERJ) for the fellowship of the Programa Cientistas do Nosso Estado, E-26/201.223/2021. L~S~R and I~A~S acknowledge the Icelandic Research Fund (Rannis), grant No. 163082-051.

\section*{Author contributions}
All authors participate in the scientific discussion of the work. All authors reviewed the paper. M.S.F., R.C.B.R., A.C.S., L.S.R., M.M., and G.L.L. edit the paper. R.C.B.R.  performed the numerical calculations. R.C.B.R., and L.S.R. performed analytical calculations.

\section*{Competing interests}
The authors declare no competing interests.

\section*{Additional information}
{\bf Supplementary Information} The online version contains supplementary material 

{\bf Correspondence} and requests for materials should be addressed to M.S.F (figueira7255@gmail.com).

\bibliographystyle{unsrt}
\bibliography{Refs}

\begin{thebibliography}{10}

\bibitem{vogt2012silicene}
P.~Vogt, P.~De Padova, C.~Quaresima, E.~Frantzeskakis Jose~Avila, M.~C.
  Asensio, A.~Resta, B.~Ealet, , and G.~{Le Lay}.
\newblock Silicene: Compelling experimental evidence for graphene like
  two-dimensional silicon.
\newblock {\em Physical Review Letters}, 108:155501, 2012.

\bibitem{Fleurence2012}
Antoine Fleurence, Rainer Friedlein, Taisuke Ozaki, Hiroyuki Kawai, Ying Wang,
  and Yukiko Yamada-Takamura.
\newblock Experimental {Evidence} for {Epitaxial} {Silicene} on {Diboride}
  {Thin} {Films}.
\newblock {\em Physical Review Letters}, 108(24):245501, June 2012.

\bibitem{Feng2012}
Baojie Feng, Zijing Ding, Sheng Meng, Yugui Yao, Xiaoyue He, Peng Cheng, Lan
  Chen, and Kehui Wu.
\newblock Evidence of silicene in honeycomb structures of silicon on ag(111).
\newblock {\em Nano Letters}, 12(7):3507--3511, 2012.
\newblock PMID: 22658061.

\bibitem{davila2022silicene}
{M.E.} Dávila and G.~Le~Lay.
\newblock Silicene: Genesis, remarkable discoveries, and legacy.
\newblock {\em Materials Today Advances}, 16:100312, 2022.

\bibitem{tao2015silicene}
L.~Tao, E.~Cinquanta, D.~Chiappe, C.~Grazianetti, M.~Fanciulli, M.~Dubey,
  A.~Molle, and D.~Akinwande.
\newblock Silicene field-effect transistors operating at room temperature.
\newblock {\em Nature Nanotechnology}, 10:227–231, 2015.

\bibitem{lelay2015silicene}
G.~Le~Lay.
\newblock Silicene transistors.
\newblock {\em Nature Nanotechnology}, 10:202, 2015.

\bibitem{frolov2013quantum}
S.~M. Frolov, S.~R. Plissard, S.~Nadj-Perge, L.~P. Kouwenhoven, and E.~P.A.M.
  Bakkers.
\newblock Quantum computing based on semiconductor nanowires.
\newblock {\em MRS Bulletin}, 38:809, 2013.

\bibitem{rancic2013exactly}
Marko~J. Rančić.
\newblock Exactly solving the kitaev chain and generating majorana-zero-modes
  out of noisy qubits.
\newblock {\em Scientic Reports}, 12:19882, 2022.

\bibitem{Flensberg2021}
Karsten Flensberg, Felix von Oppen, and Ady Stern.
\newblock Engineered platforms for topological superconductivity and majorana
  zero modes.
\newblock {\em Nature Reviews Materials}, Jul 2021.

\bibitem{sheng2018pentagonal}
S.~Sheng, R.~Ma, J.~b.~Wu, W.~Li, L.~Kong, X.~Cong, D.~Cao, W.~Hu, J.~Gou,
  J.-W. Luo, P.~Cheng, P.-H. Tan, Y.~Jiang, L.~Chen, and K.~Wu.
\newblock The pentagonal nature of self-assembled silicon chains and magic
  clusters on ag(110).
\newblock {\em Nano Lett.}, 18:2937, 2018.

\bibitem{cerda2016unveiling}
Jorge~I Cerd{\'a}, Jagoda S{\l}awi{\'n}ska, Guy Le~Lay, Antonela~C Marele,
  Jos{\'e}~M G{\'o}mez-Rodr{\'\i}guez, and Mar{\'\i}a D{\'a}vila.
\newblock Unveiling the pentagonal nature of perfectly aligned single-and
  double-strand si nano-ribbons on ag(110).
\newblock {\em Nature Communications}, 7:13076, 2016.

\bibitem{Kitaev2001}
A~Yu Kitaev.
\newblock Unpaired majorana fermions in quantum wires.
\newblock {\em Physics-Uspekhi}, 44(10S):131, 2001.

\bibitem{Oreg2010}
Yuval Oreg, Gil Refael, and Felix von Oppen.
\newblock Helical liquids and majorana bound states in quantum wires.
\newblock {\em Phys. Rev. Lett.}, 105:177002, Oct 2010.

\bibitem{Aguado17}
R\'amon Aguado.
\newblock Majorana quasiparticles in condensed matter.
\newblock {\em Riv Nuovo Cimento}, 40(11):523, 2017.

\bibitem{Schuray2017}
Alexander Schuray, Luzie Weithofer, and Patrik Recher.
\newblock Fano resonances in majorana bound states--quantum dot hybrid systems.
\newblock {\em Phys. Rev. B}, 96:085417, Aug 2017.

\bibitem{Prada2018}
Elsa Prada, Ram\'on Aguado, and Pablo San-Jose.
\newblock Measuring majorana nonlocality and spin structure with a quantum dot.
\newblock {\em Phys. Rev. B}, 96:085418, Aug 2017.

\bibitem{Ricco2018}
L.~S. Ricco, M.~de~Souza, M.~S. Figueira, I.~A. Shelykh, and A.~C. Seridonio.
\newblock Spin-dependent zero-bias peak in a hybrid nanowire-quantum dot
  system: Distinguishing isolated majorana fermions from andreev bound states.
\newblock {\em Phys. Rev. B}, 99:155159, Apr 2019.

\bibitem{ZhangNatCommun2019}
Hao Zhang, Dong~E. Liu, Michael Wimmer, and Leo~P. Kouwenhoven.
\newblock Next steps of quantum transport in majorana nanowire devices.
\newblock {\em Nat Commun}, 10:5128, Nov 2019.

\bibitem{prada2020andreev}
Elsa Prada, Pablo San-Jose, Michiel W.~A. de~Moor, Attila Geresdi, Eduardo
  J.~H. Lee, Jelena Klinovaja, Daniel Loss, Jesper Nygård, Ramón Aguado, and
  Leo~P. Kouwenhoven.
\newblock From andreev to majorana bound states in hybrid
  superconductor-semiconductor nanowires.
\newblock {\em Nat Rev Phys}, 2:575--594, 2020.

\bibitem{Oreg2012}
Anindya Das, Yuval Ronen, Yonatan Most, Yuval Oreg, Moty Heiblum, and Hadas
  Shtrikman.
\newblock Zero-bias peaks and splitting in an al-inas nanowire topological
  superconductor as a signature of majorana fermions.
\newblock {\em Nature Physics}, 8:887 EP --, Nov 2012.
\newblock Article.

\bibitem{Mourik12}
V.~Mourik, K.~Zuo, S.~M. Frolov, S.~R. Plissard, E.~P. A.~M. Bakkers, and L.~P.
  Kouwenhoven.
\newblock Signatures of majorana fermions in hybrid
  superconductor-semiconductor nanowire devices.
\newblock {\em Science}, 336(6084):1003--1007, 2012.

\bibitem{Perge14}
Stevan Nadj-Perge, Ilya~K. Drozdov, Jian Li, Hua Chen, Sangjun Jeon, Jungpil
  Seo, Allan~H. MacDonald, B.~Andrei Bernevig, and Ali Yazdani.
\newblock Observation of majorana fermions in ferromagnetic atomic chains on a
  superconductor.
\newblock {\em Science}, 346(6209):602--607, 2014.

\bibitem{Krogstrup2015}
P.~Krogstrup, N.~L.~B. Ziino, W.~Chang, S.~M. Albrecht, M.~H. Madsen,
  E.~Johnson, J.~Nyg{\aa}rd, C.~M. Marcus, and T.~S. Jespersen.
\newblock Epitaxy of semiconductor--superconductor nanowires.
\newblock {\em Nature Materials}, 14(4):400--406, Apr 2015.

\bibitem{JeonScience2017}
Sangjun Jeon, Yonglong Xie, Jian Li, Zhijun Wang, B.~Andrei Bernevig, and Ali
  Yazdani.
\newblock Distinguishing a majorana zero mode using spin-resolved measurements.
\newblock {\em Science}, 358(6364):772--776, 2017.

\bibitem{Clarke2017}
David~J. Clarke.
\newblock Experimentally accessible topological quality factor for wires with
  zero energy modes.
\newblock {\em Phys. Rev. B}, 96:201109(R), Nov 2017.

\bibitem{Gul2018}
{\"O}nder G{\"u}l, Hao Zhang, Jouri D.~S. Bommer, Michiel W.~A. de~Moor, Diana
  Car, S{\'e}bastien~R. Plissard, Erik P. A.~M. Bakkers, Attila Geresdi, Kenji
  Watanabe, Takashi Taniguchi, and Leo~P. Kouwenhoven.
\newblock Ballistic majorana nanowire devices.
\newblock {\em Nature Nanotechnology}, 13(3):192--197, Mar 2018.

\bibitem{LutchynReviewMat2018}
R.~M. Lutchyn, E.~P. A.~M. Bakkers, L.~P. Kouwenhoven, P.~Krogstrup, C.~M.
  Marcus, and Y.~Oreg.
\newblock Majorana zero modes in superconductor–semiconductor
  heterostructures.
\newblock {\em Nature Reviews Materials}, 3:52--68, May 2018.

\bibitem{jack2021}
B.~J\"{a}ck, Y.~Xie, and A.~Yazdani.
\newblock {\em Nature Reviews Physics}, 3:541, 2021.

\bibitem{Pan2020}
Haining Pan and S.~Das~Sarma.
\newblock Physical mechanisms for zero-bias conductance peaks in majorana
  nanowires.
\newblock {\em Phys. Rev. Research}, 2:013377, Mar 2020.

\bibitem{PanGenericQuantized2020}
Haining Pan, William~S. Cole, Jay~D. Sau, and S.~Das~Sarma.
\newblock Generic quantized zero-bias conductance peaks in
  superconductor-semiconductor hybrid structures.
\newblock {\em Phys. Rev. B}, 101:024506, Jan 2020.

\bibitem{pan2021quantized}
Haining Pan, Chun-Xiao Liu, Michael Wimmer, and Sankar Das~Sarma.
\newblock Quantized and unquantized zero-bias tunneling conductance peaks in
  majorana nanowires: Conductance below and above $2{e}^{2}/h$.
\newblock {\em Phys. Rev. B}, 103:214502, Jun 2021.

\bibitem{Kim2018}
Howon Kim, Alexandra Palacio-Morales, Thore Posske, Levente Rózsa, Krisztián
  Palotás, László Szunyogh, Michael Thorwart, and Roland Wiesendanger.
\newblock Toward tailoring majorana bound states in artificially constructed
  magnetic atom chains on elemental superconductors.
\newblock {\em Science Advances}, 4(5):eaar5251, 2018.

\bibitem{Jelena_bilayer_graphene_2012}
Jelena Klinovaja, Gerson~J. Ferreira, and Daniel Loss.
\newblock Helical states in curved bilayer graphene.
\newblock {\em Phys. Rev. B}, 86:235416, Dec 2012.

\bibitem{Jelena_nanoribbons_2013}
Jelena Klinovaja and Daniel Loss.
\newblock Giant spin-orbit interaction due to rotating magnetic fields in
  graphene nanoribbons.
\newblock {\em Phys. Rev. X}, 3:011008, Jan 2013.

\bibitem{Maiellaro2018}
Alfonso Maiellaro, Francesco Romeo, and Roberta Citro.
\newblock Topological phase diagram of a kitaev ladder.
\newblock {\em The European Physical Journal Special Topics},
  227(12):1397--1404, Dec 2018.

\bibitem{Bento2022}
R.~C.~Bento Ribeiro, J.~H. Correa, L.~S. Ricco, A.~C. Seridonio, and M.~S.
  Figueira.
\newblock Spin-polarized majorana zero modes in double zigzag honeycomb
  nanoribbons.
\newblock {\em Phys. Rev. B}, 105:205115, May 2022.

\bibitem{Dutreix2014}
Cl{\'e}ment Dutreix, Marine Guigou, Denis Chevallier, and Cristina Bena.
\newblock Majorana fermions in honeycomb lattices.
\newblock {\em The European Physical Journal B}, 87(12):296, Dec 2014.

\bibitem{Ma2017}
Tianxing Ma, Fan Yang, Zhongbing Huang, and Hai-Qing Lin.
\newblock Triplet p-wave pairing correlation in low-doped zigzag graphene
  nanoribbons.
\newblock {\em Scientific Reports}, 7:42262 EP --, Feb 2017.
\newblock Article.

\bibitem{Aidi20}
Aidi Zhaoa and Bing Wang.
\newblock Two-dimensional graphene-like xenes as potential topological
  materials.
\newblock {\em APL Materials}, 8:030701, 2020.

\bibitem{Grazianetti2021}
Carlo Grazianetti and Christian Martella.
\newblock The rise of the xenes: From the synthesis to the integration
  processes for electronics and photonics.
\newblock {\em Materials (Basel)}, 14(15), July 2021.

\bibitem{yue2022}
S.~Yue, H.~Zhou, Y.~Feng, Y.~Wang, Z.~Sun, D.~Geng, M.~Arita, S.~Kumar,
  K.~Shimada, P.~Cheng, L.~Chen, Y.~Yao, S.~Meng, K.~Wu, , and B.~Feng.
\newblock {\em Nano Lett.}, 22:695, 2022.

\bibitem{Depadova2010}
Paola De~Padova, Claudio Quaresima, Carlo Ottaviani, Polina~M. Sheverdyaeva,
  Paolo Moras, Carlo Carbone, Dinesh Topwal, Bruno Olivieri, Abdelkader Kara,
  Hamid Oughaddou, Bernard Aufray, and Guy Le~Lay.
\newblock Evidence of graphene-like electronic signature in silicene
  nanoribbons.
\newblock {\em Applied Physics Letters}, 96(26), June 2010.

\bibitem{DePadova2012}
Paola~De Padova, Paolo Perfetti, Bruno Olivieri, Claudio Quaresima, Carlo
  Ottaviani, and Guy~Le Lay.
\newblock 1d graphene-like silicon systems: silicene nano-ribbons.
\newblock {\em Journal of Physics: Condensed Matter}, 24(22):223001, may 2012.

\bibitem{Iribas2019}
R{\'e}my Pawlak, Carl Drechsel, Philipp D’Astolfo, Marcin Kisiel, Ernst
  Meyer, and Jorge~Iribas Cerda.
\newblock Quantitative determination of atomic buckling of silicene by atomic
  force microscopy.
\newblock {\em Proceedings of the National Academy of Sciences},
  117(1):228--237, 2020.

\bibitem{Prince2003}
N~Tsud, S~Fabik, V~Dudr, M~Vondracek, V~Chab, V~Matolin, and K.C Prince.
\newblock Interfacial reconstruction in the system pb/ag(110).
\newblock {\em Surface Science}, 542(1):112--119, 2003.

\bibitem{Krawiec2015}
Agata Podsiadły-Paszkowska and Mariusz Krawiec.
\newblock Dirac fermions in silicene on pb(111) surface.
\newblock {\em Phys. Chem. Chem. Phys.}, 17:2246--2251, 2015.

\bibitem{Krawiec2019}
Agnieszka Stepniak-Dybala and Mariusz Krawiec.
\newblock Formation of silicene on ultrathin pb(111) films.
\newblock {\em The Journal of Physical Chemistry C}, 123(27):17019--17025, Jul
  2019.

\bibitem{Yazdani2021}
Berthold J{\"a}ck, Yonglong Xie, and Ali Yazdani.
\newblock Detecting and distinguishing majorana zero modes with the scanning
  tunnelling microscope.
\newblock {\em Nature Reviews Physics}, 3(8):541--554, Aug 2021.

\bibitem{Min2006}
Hongki Min, J.~E. Hill, N.~A. Sinitsyn, B.~R. Sahu, Leonard Kleinman, and A.~H.
  MacDonald.
\newblock Intrinsic and rashba spin-orbit interactions in graphene sheets.
\newblock {\em Phys. Rev. B}, 74:165310, Oct 2006.

\bibitem{zarea2009}
Mahdi Zarea and Nancy Sandler.
\newblock Rashba spin-orbit interaction in graphene and zigzag nanoribbons.
\newblock {\em Phys. Rev. B}, 79:165442, Apr 2009.

\bibitem{Ezawa2012}
Motohiko Ezawa.
\newblock Valley-polarized metals and quantum anomalous hall effect in
  silicene.
\newblock {\em Phys. Rev. Lett.}, 109:055502, Aug 2012.

\bibitem{Jiao2020}
Z.~Jiao, Q.~Yao, and H.J.W. Zandvliet.
\newblock Tailoring and probing the quantum states of matter of $2d$ dirac
  materials with a buckled honeycomb structure.
\newblock {\em Physica E: Low-dimensional Systems and Nanostructures},
  121:114113, 2020.

\bibitem{Alase2019}
Abhijeet Alase.
\newblock {\em Boundary Physics and Bulk-Boundary Correspondence in Topological
  Phases of Matter}.
\newblock Springer Theses, Springer Nature Switzerland AG, 1 edition, 2019.

\bibitem{ZakPhysRevLett.62.2747(1989)}
J.~Zak.
\newblock Berry's phase for energy bands in solids.
\newblock {\em Phys. Rev. Lett.}, 62:2747--2750, Jun 1989.

\bibitem{Tsai2013}
Wei-Feng Tsai, Cheng-Yi Huang, Tay-Rong Chang, Hsin Lin, Horng-Tay Jeng, and
  A.~Bansil.
\newblock Gated silicene as a tunable source of nearly 100{\%} spin-polarized
  electrons.
\newblock {\em Nature Communications}, 4(1):1500, Feb 2013.

\bibitem{Jiang_2019}
Peng Jiang, Lili Kang, Xixi Tao, Ning Cao, Hua Hao, Xiaohong Zheng, Lei Zhang,
  and Zhi Zeng.
\newblock Robust generation of half-metallic transport and pure spin current
  with photogalvanic effect in zigzag silicene nanoribbons.
\newblock {\em Journal of Physics: Condensed Matter}, 31(49):495701, sep 2019.

\bibitem{Ensslin2010}
T.~Ihn, J.~Güttinger, F.~Molitor, S.~Schnez, E.~Schurtenberger, A.~Jacobsen,
  S.~Hellmüller, T.~Frey, S.~Dröscher, C.~Stampfer, and K.~Ensslin.
\newblock Graphene single-electron transistors.
\newblock {\em Materials Today}, 13(3):44--50, 2010.

\bibitem{Niu2023}
Wenhui Niu, Simen Sopp, Alessandro Lodi, Alex Gee, Fanmiao Kong, Tian Pei,
  Pascal Gehring, Jonathan N{\"a}gele, Chit~Siong Lau, Ji~Ma, Junzhi Liu,
  Akimitsu Narita, Jan Mol, Marko Burghard, Klaus M{\"u}llen, Yiyong Mai,
  Xinliang Feng, and Lapo Bogani.
\newblock Exceptionally clean single-electron transistors from solutions of
  molecular graphene nanoribbons.
\newblock {\em Nature Materials}, 22(2):180--185, Feb 2023.

\bibitem{Goldhaber1998}
D.~Goldhaber-Gordon, Hadas Shtrikman, D.~Mahalu, David Abusch-Magder,
  U.~Meirav, and M.~A. Kastner.
\newblock Kondo effect in a single-electron transistor.
\newblock {\em Nature}, 391(6663):156--159, Jan 1998.

\bibitem{VernekLeakagePhysRevB.89.165314(2015)}
E.~Vernek, P.~H. Penteado, A.~C. Seridonio, and J.~C. Egues.
\newblock Subtle leakage of a majorana mode into a quantum dot.
\newblock {\em Phys. Rev. B}, 89:165314, Apr 2014.

\bibitem{KarzigPRB2017}
Torsten Karzig, Christina Knapp, Roman~M. Lutchyn, Parsa Bonderson, Matthew~B.
  Hastings, Chetan Nayak, Jason Alicea, Karsten Flensberg, Stephan Plugge,
  Yuval Oreg, Charles~M. Marcus, and Michael~H. Freedman.
\newblock Scalable designs for quasiparticle-poisoning-protected topological
  quantum computation with majorana zero modes.
\newblock {\em Phys. Rev. B}, 95:235305, Jun 2017.

\bibitem{Steiner2020}
Jacob~F. Steiner and Felix von Oppen.
\newblock Readout of majorana qubits.
\newblock {\em Phys. Rev. Research}, 2:033255, Aug 2020.

\bibitem{LeijnsePRL2011}
Martin Leijnse and Karsten Flensberg.
\newblock Quantum information transfer between topological and spin qubit
  systems.
\newblock {\em Phys. Rev. Lett.}, 107:210502, Nov 2011.

\bibitem{Flensberg2012}
Martin Leijnse and Karsten Flensberg.
\newblock Hybrid topological-spin qubit systems for two-qubit-spin gates.
\newblock {\em Phys. Rev. B}, 86:104511, Sep 2012.

\bibitem{Sarma2016}
Xin Liu, Xiaopeng Li, Dong-Ling Deng, Xiong-Jun Liu, and S.~Das~Sarma.
\newblock Majorana spintronics.
\newblock {\em Phys. Rev. B}, 94:014511, Jul 2016.

\bibitem{BernevigBook}
B.~Andrei Bernevig and Taylor~L. Hughes.
\newblock {\em {Topological Insulators and Topological Superconductors}}.
\newblock {Princeton University Press}, {STU - Student edition} edition, 2013.

\bibitem{Chiu2016}
Ching-Kai Chiu, Jeffrey C.~Y. Teo, Andreas~P. Schnyder, and Shinsei Ryu.
\newblock Classification of topological quantum matter with symmetries.
\newblock {\em Rev. Mod. Phys.}, 88:035005, Aug 2016.

\bibitem{StanescuBook}
Tudor~D. Stanescu.
\newblock {\em {Introduction to Topological Quantum Matter \& Quantum
  Computation}}.
\newblock {CRC Press}, 1st edition edition, 2016.

\end{thebibliography}

\end{document}